\documentclass[11pt,twoside]{article}

\usepackage{asp2004}
\usepackage{graphicx}

\begin{document}
\title{The Hanle and Zeeman Effects in Solar Spicules}   
\author{R. Ramelli, M. Bianda}   
\affil{Istituto Ricerche Solari Locarno, CH-6605 Locarno Monti, Switzerland} 
\author{L. Merenda, J. Trujillo Bueno}
\affil{Instituto de Astrof\'isica de Canarias, E-38205 La Laguna, Tenerife, Spain}

\begin{abstract} 
A large set of high precision full-Stokes spectropolarimetric
observations of the He-D$_3$ line in spicules has been  
recorded with the ZIMPOL polarimeter at the Gregory-Coud\'e Telescope in Locarno.
The observational technique allow us to obtain measurements
free from seeing induced spurious effects.
The instrumental polarization is well under control and taken
into account in the data analysis. The observed Stokes profiles
are interpreted according to the quantum theory of
the Hanle and Zeeman effects with the aim of obtaining
information on the magnetic field vector. To this end, we
make use of a suitable Stokes inversion technique.
The results are presented giving emphasis on a few particularly
interesting measurements which show clearly the operation of the  
Hanle and Zeeman effects in solar chromospheric spicules.
\end{abstract}

\section{Introduction}
The mechanism governing the formation of spicules is still
uncertain. None of the theoretical models proposed until now
\citep[see e.g. the review by][]{Sterling2000} 
is able to explain in a satisfactory way
all observational facts including ubiquity, evolution, energetics and
periodicity \citep{Depontieu2004} and also the magnetic fields,
whose presence in spicules is required by all theoretical models.
Investigations of
such magnetic fields through spectropolarimetric observation were
reported recently \citep{trujillo05,lopez05a,ramellilindau05}.
Together with a suitable inversion technique,
which takes into account the joint action of the Hanle and Zeeman
effects, full-Stokes polarimetry is in fact 
a powerful diagnostic tool to obtain information about the
three-dimensional geometry of the magnetic fields that channel the 
spicular motion.
In particular the Helium multiplets at 5876 and 10830~\AA~are very useful spectral lines for that purpose. 

In the past year an ambitious observational project of spicules 
in the Helium D$_3$ multiplet has been stared at the Istituto Ricerche Solari
Locarno (IRSOL) taking advantage of the
good performances
of the Zurich Imaging Polarimeter (ZIMPOL) \citep{gandorfer04}.
A preliminary physical interpretation of our observations based on suitable 
inversion techniques has been applied in order
to obtain information on the magnetic field vectors involved.

With the same technique we have additionally 
obtained several prominence
observations, which are described also in these proceedings
\citep{Ramelliprom}.

\section{The Observations}

The observations of spicules were performed with the Gregory-Coud\'e
Telescope (GCT) at IRSOL, whose aperture is 45 cm.
The ZIMPOL polarimeter \citep{gandorfer04} allowed precise measurements free from seeing 
induced spurious effects (modulation at 42 kHz). 
The solar image was rotated with a Dove prism set after the polarization
analyzer in order to keep the limb parallel to the spectrograph slit. 
A limb tracker was used to maintain the distance from the limb constant.
53 spicules measurements were obtained at different latitudes and limb
distances during 15 days in the period between November 2004 and June
2005. 
The total exposure time for each measurement ranged from 10 minutes to
about 1 hour. Every 2 minutes simultaneous measurements of Stokes $I$, $V/I$ and one linear polarization
component (alternatively $Q/I$ and $U/I$) was stored allowing to observe
the time evolution. Despite of the
fact that the dynamic jets that we call spicules
have a lifetime of 5-10 minutes, the
general structures of the obtained Stokes images remained almost similar even during the long measurements.

Calibrations observations were performed regularly. These included 
polarimetric efficiency measurements, dark current, flat field and measurements
of the instrumental polarization. 

\begin{figure}[b!]
\includegraphics[angle=0,width=.47\linewidth]{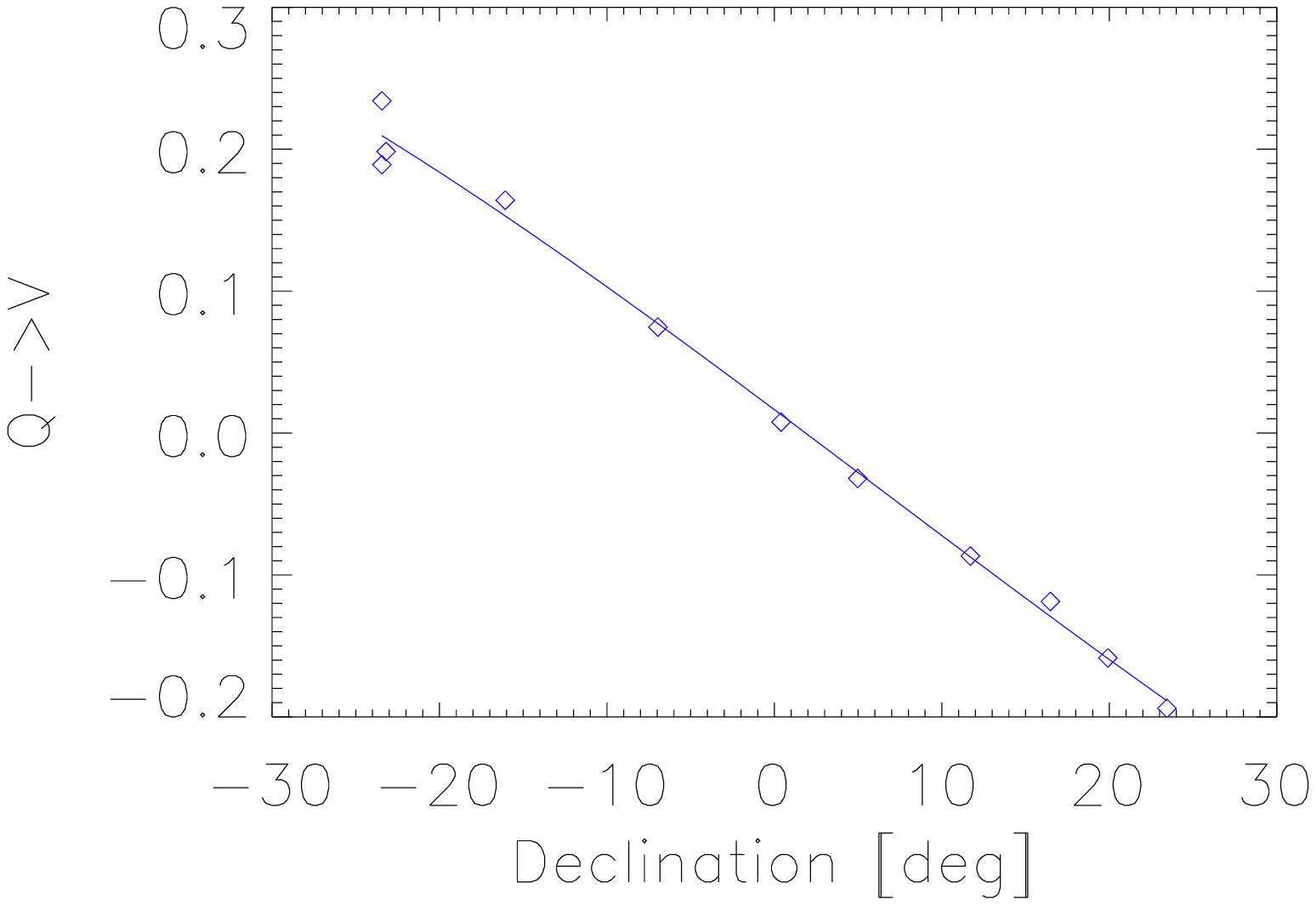}
\hfill
\includegraphics[angle=0,width=.47\linewidth]{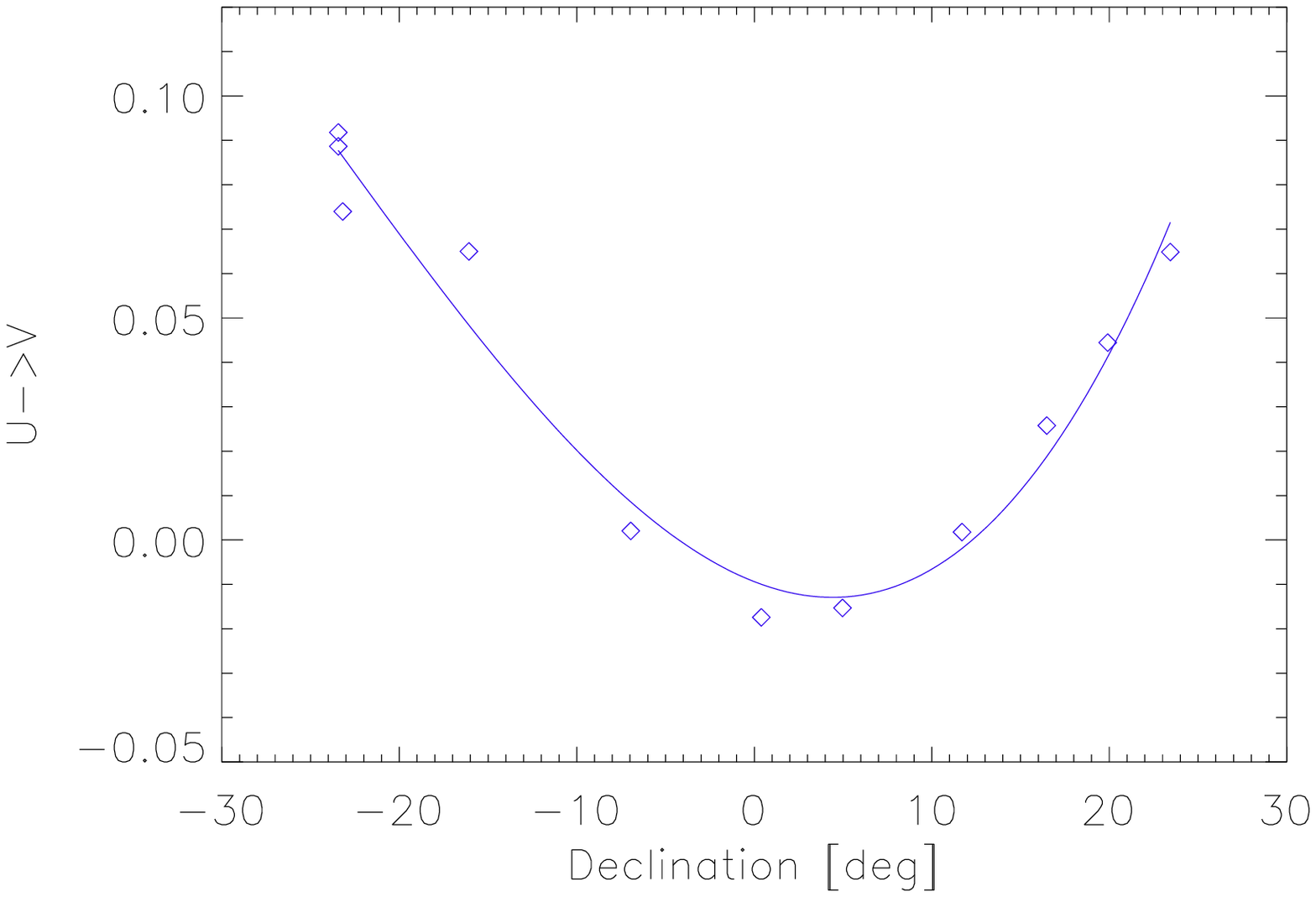}
\caption{\label{fitctallwl-qv}{ Measurement of linear to circular
crosstalks  as a function of declination for He-D$_3$. {\itshape Left:\/} $Q\rightarrow V$ crosstalk.
To obtain this measurement a linear polarizer is placed
before the entrance window of the telescope with the polarization
axis parallel to the geographic equator, which is our choice here for the
positive direction of Stokes $Q$.} 
{\itshape Right:\/}  $U\rightarrow V$ crosstalk. 
To obtain this measurement the linear polarizer is rotated by
45$^\circ$ with respect to the previous measurement. }
\end{figure}

The instrumental polarization could be carefully corrected for, taking
advantage of the fact that in a GCT the instrumental polarization for a given wavelength is a function of 
declination and stays therefore 
almost constant over one day. 
The crosstalks $I\rightarrow Q$, $I\rightarrow U$ and $I\rightarrow V$ were determined through 
the measurements performed in quiet regions near the solar disk center.
The crosstalks $Q\rightarrow V$ and $U\rightarrow V$ were determined applying
a linear polarization filter at different positions before the entrance window 
of the telescope. The results as a function of the declination are shown
in Fig.~\ref{fitctallwl-qv}.
The $V\rightarrow Q$ and $V\rightarrow U$ crosstalks were extracted from the $Q\rightarrow V$ and $U\rightarrow V$ 
crosstalk measurements taking into account the symmetries of the 
theoretical M\"uller matrix of a GCT \citep[see][]{sanchez91}. 

\section{Inference of the magnetic field from the He-D$_3$ profiles}
A database containing the theoretical Stokes profiles for
different limb distances, magnetic field orientations and strengths
has been created, with the theoretical Stokes profiles calculated via
the application of the quantum theory of the Hanle and 
Zeeman effects \citep[see, e.g.,][]{innocenti04}
Spicules are assumed to be optically thin. 
The theoretical profiles that better fit the measured profiles are carefully
searched in the database, in order to infer the magnetic field vector. 

Future planed improvements in our analysis will include the introduction of
non-thermal motions to better fit the intensity profiles and a detailed study
of the so-called Van Vleck
ambiguity which may occur in addition to the well known 180$^\circ$
ambiguity. We will eventually evaluate the opportunity to 
account for radiative transfer effects \citep[see][]{trujillo05}, although they are usually small for the He-D$_3$ line.

\section{Results}

\begin{figure}[h!]
\includegraphics[angle=0,width=.47\linewidth]{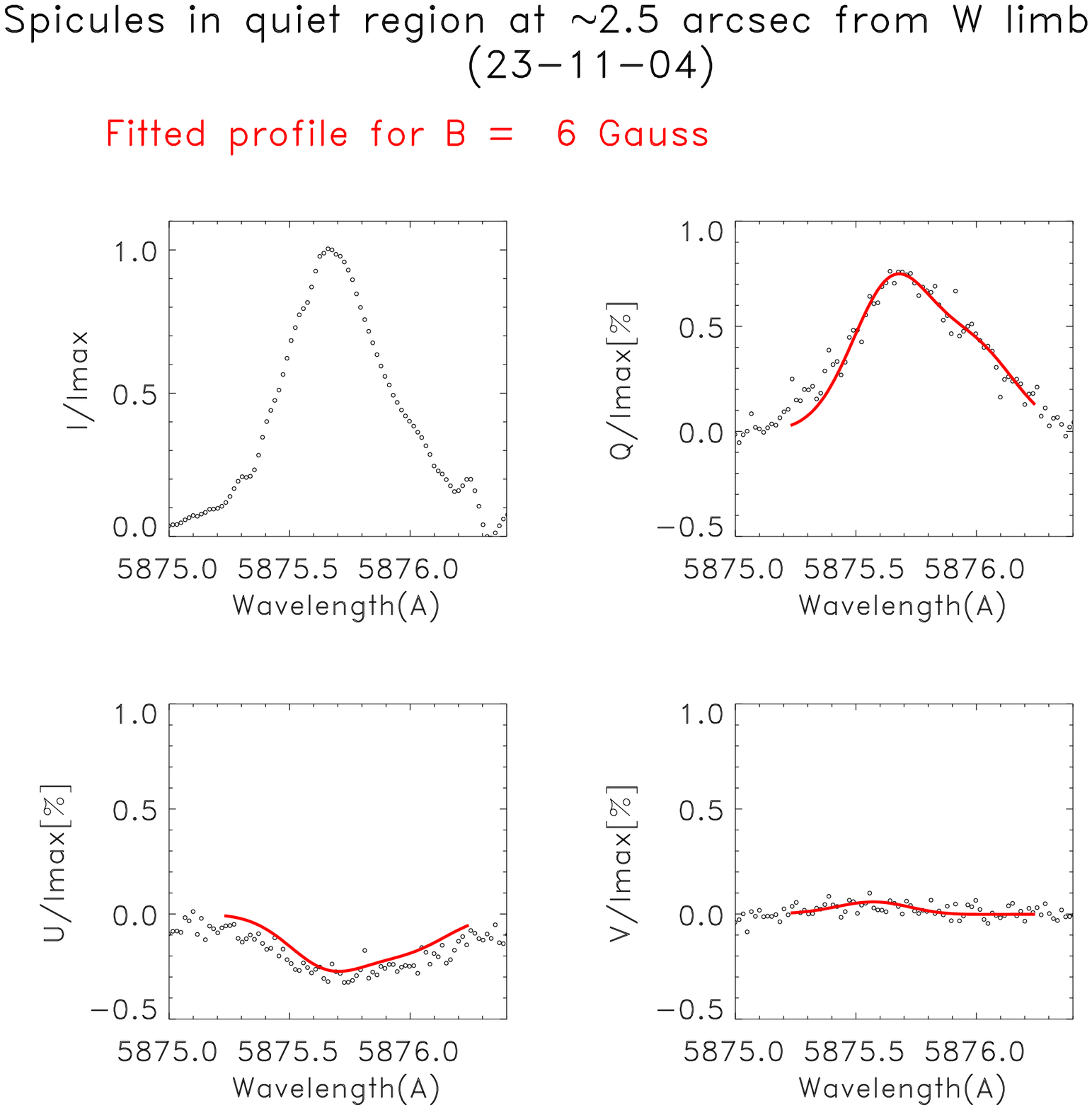}
\hfill
\includegraphics[angle=0,width=.47\linewidth]{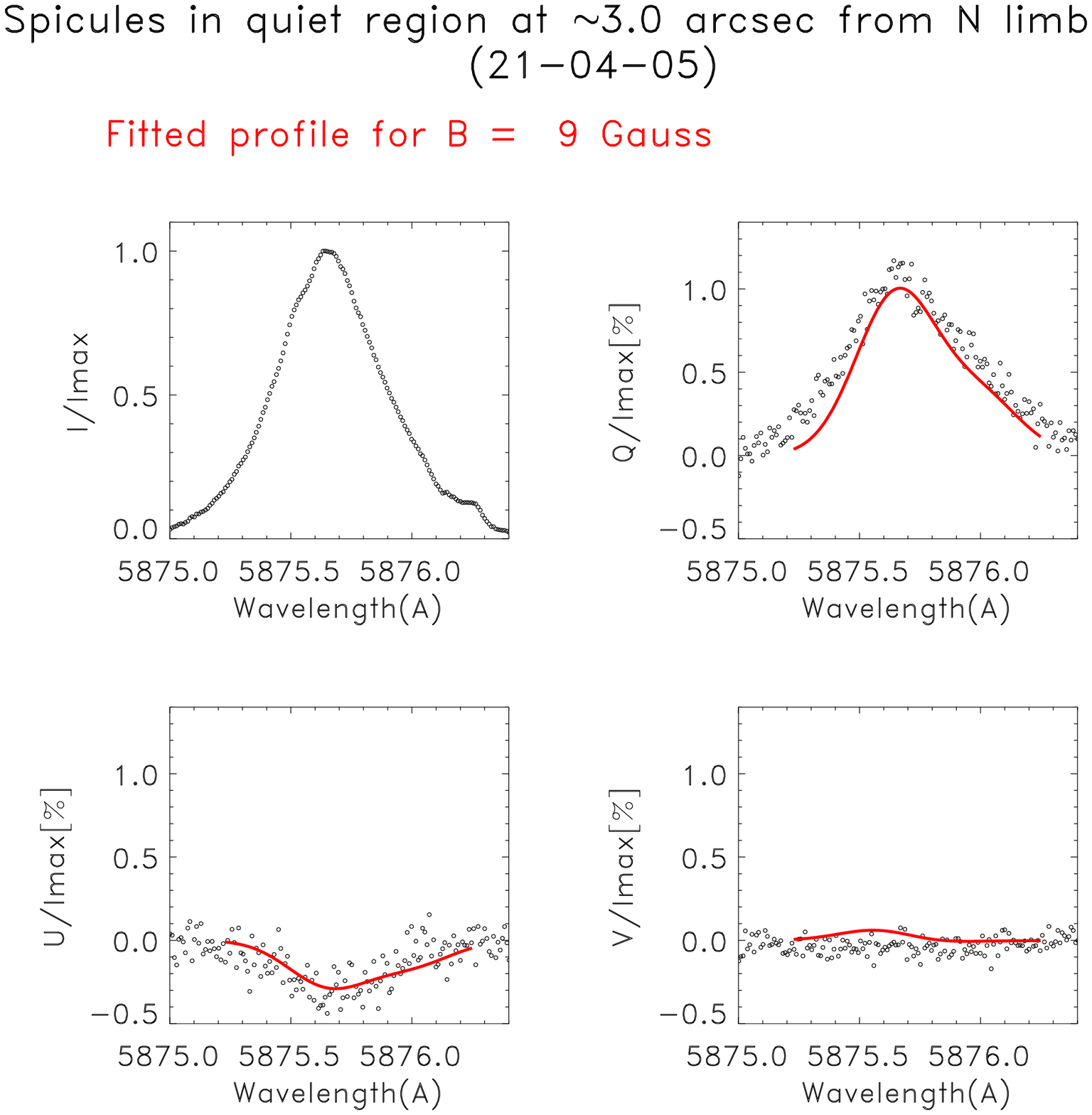}
\caption{Examples of Stokes profiles observed 
in a quiet region obtained after
integrating 60 arcseconds along the direction specified by the 
spectrograph slit.
The solid lines show the corresponding theoretical
fits.\label{prof-quiet02}}
\end{figure}

\begin{figure}[!h]
\begin{center}
\includegraphics[angle=90,width=0.85\linewidth]{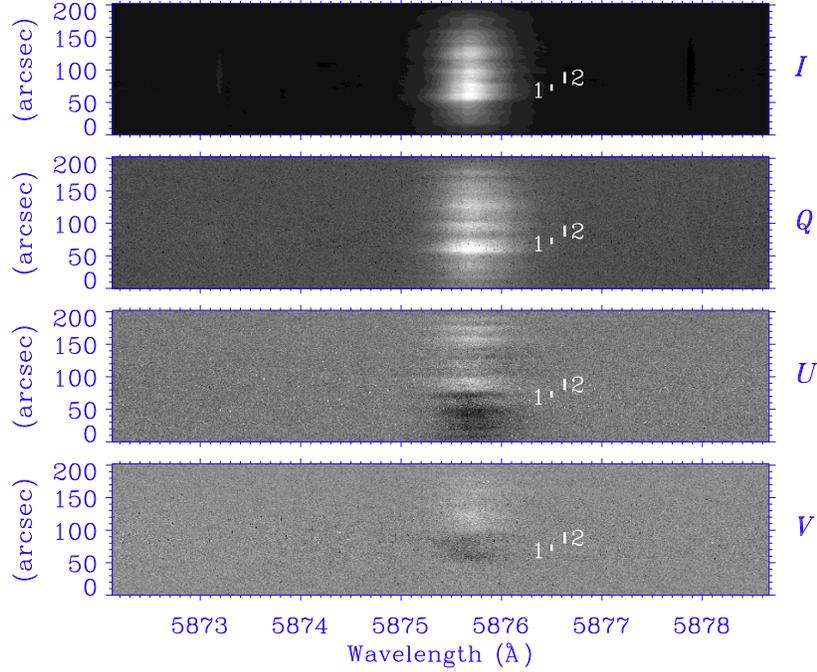}
\caption{A measurement of spicules in a quiet region.
Note that the $V$-profile is dominated by the alignment-to-orientation
conversion mechanism explained in \citet{kemp84} and
\citet{innocenti04}.
Two regions marked with ``1'' and ``2'' are selected for inversions, which are shown
in the next figure.\label{5876-050618sp5876m2}}
\end{center}
\end{figure}

\begin{figure}[!h]
\includegraphics[angle=0,width=.47\linewidth]{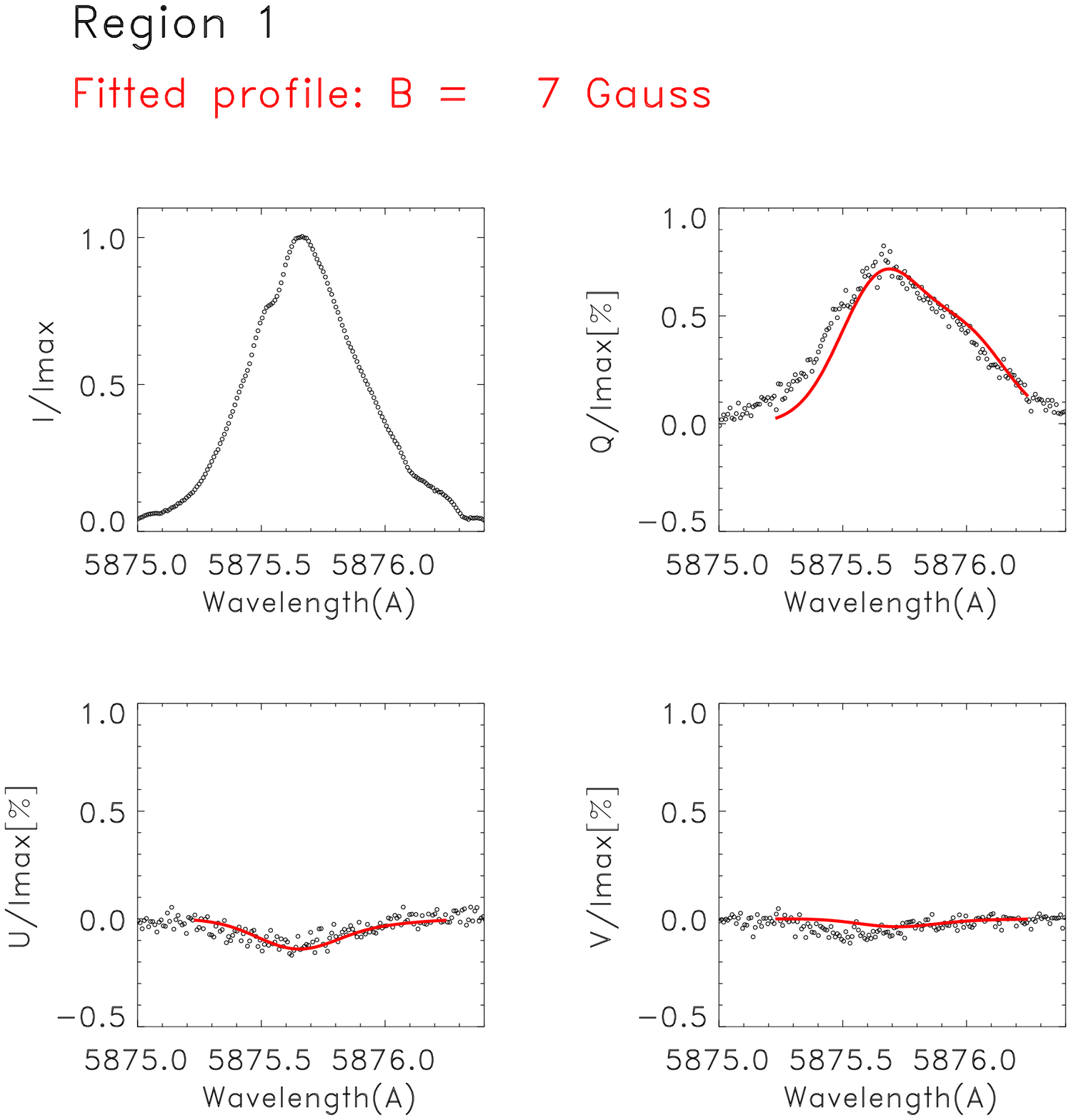}
\hfill
\includegraphics[angle=0,width=.47\linewidth]{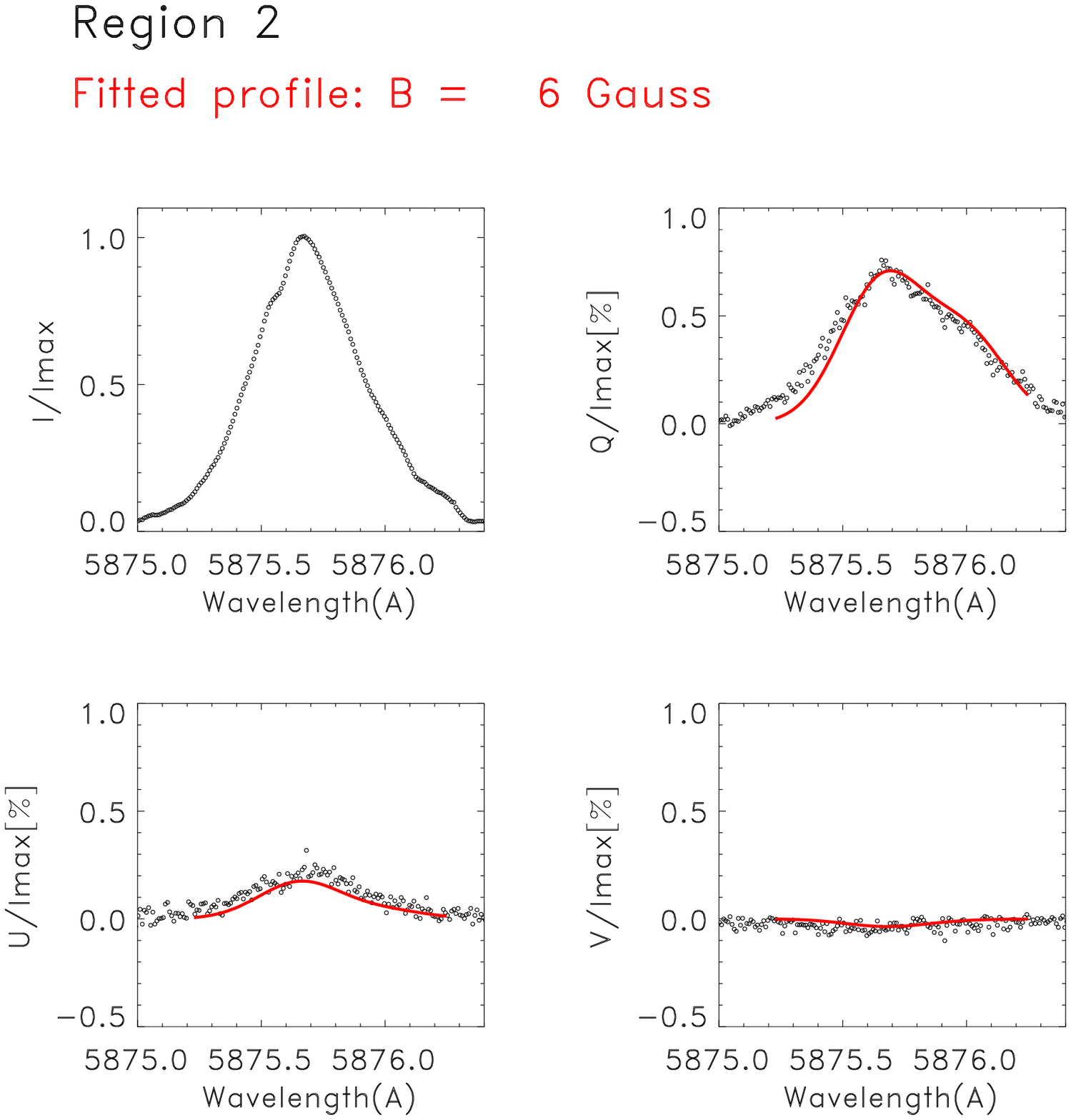}
\caption{\label{prof-NW}Stokes profiles corresponding to the
two regions selected in Figure \ref{5876-050618sp5876m2} with the
ensuing theoretical fits (continuous lines)
}
\end{figure}

A preliminary analysis of our observations in quiet
regions shows that generally the measured
He-D$_3$ Stokes profiles are compatible with a magnetic field strength of 
approximately 10
gauss (see examples in Fig.~\ref{prof-quiet02},
\ref{5876-050618sp5876m2} and \ref{prof-NW}), which is in good agreement
with the results obtained from the He {\sc i} 10830~\AA~multiplet by
\citet{trujillo05}. The Stokes $V$ profiles are usually
dominated by the symmetric signature 
due to the alignment-to-orientation transfer mechanism discussed by
\citet{kemp84} and \citet{innocenti04}.

\begin{figure}[h!]
\begin{center}
\includegraphics[angle=90,width=0.8\linewidth]{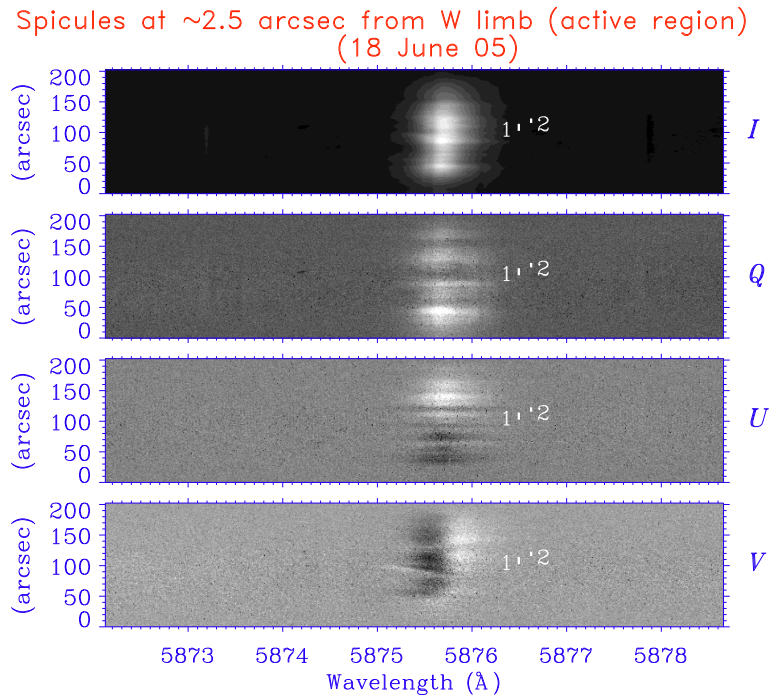}
\caption{Spectropolarimetric observation
of spicules near an active region showing antisymmetric
Zeeman-like Stokes $V$ profiles, which indicates stronger magnetic fields.
Two regions marked with ``1'' and ``2'' are selected for inversions, which are shown
in the next figure.\label{5876-050618sp5876m4}}
\end{center}
\end{figure}

\begin{figure}[h!]
\includegraphics[angle=0,width=.47\linewidth]{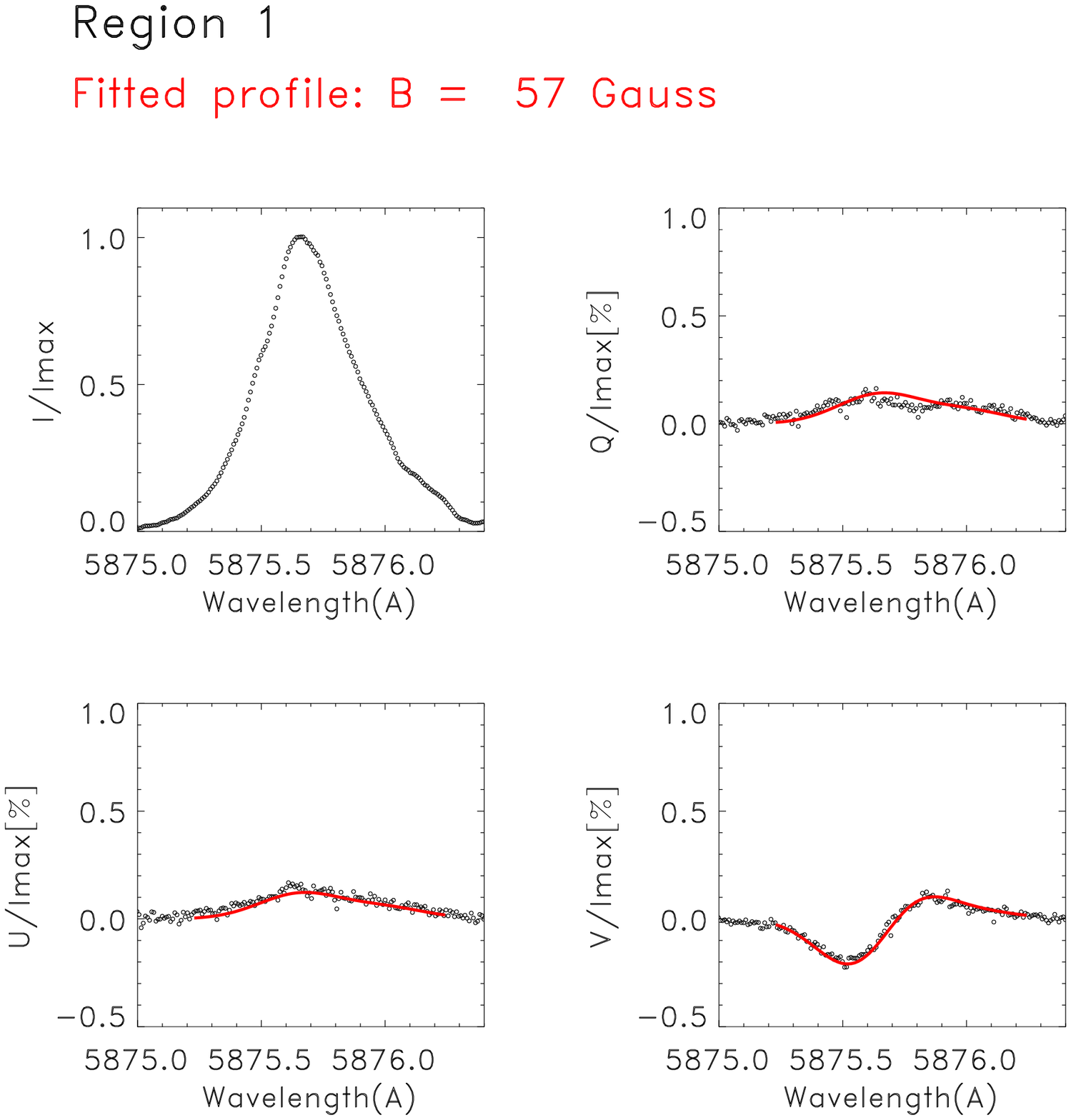}
\hfill
\includegraphics[angle=0,width=.47\linewidth]{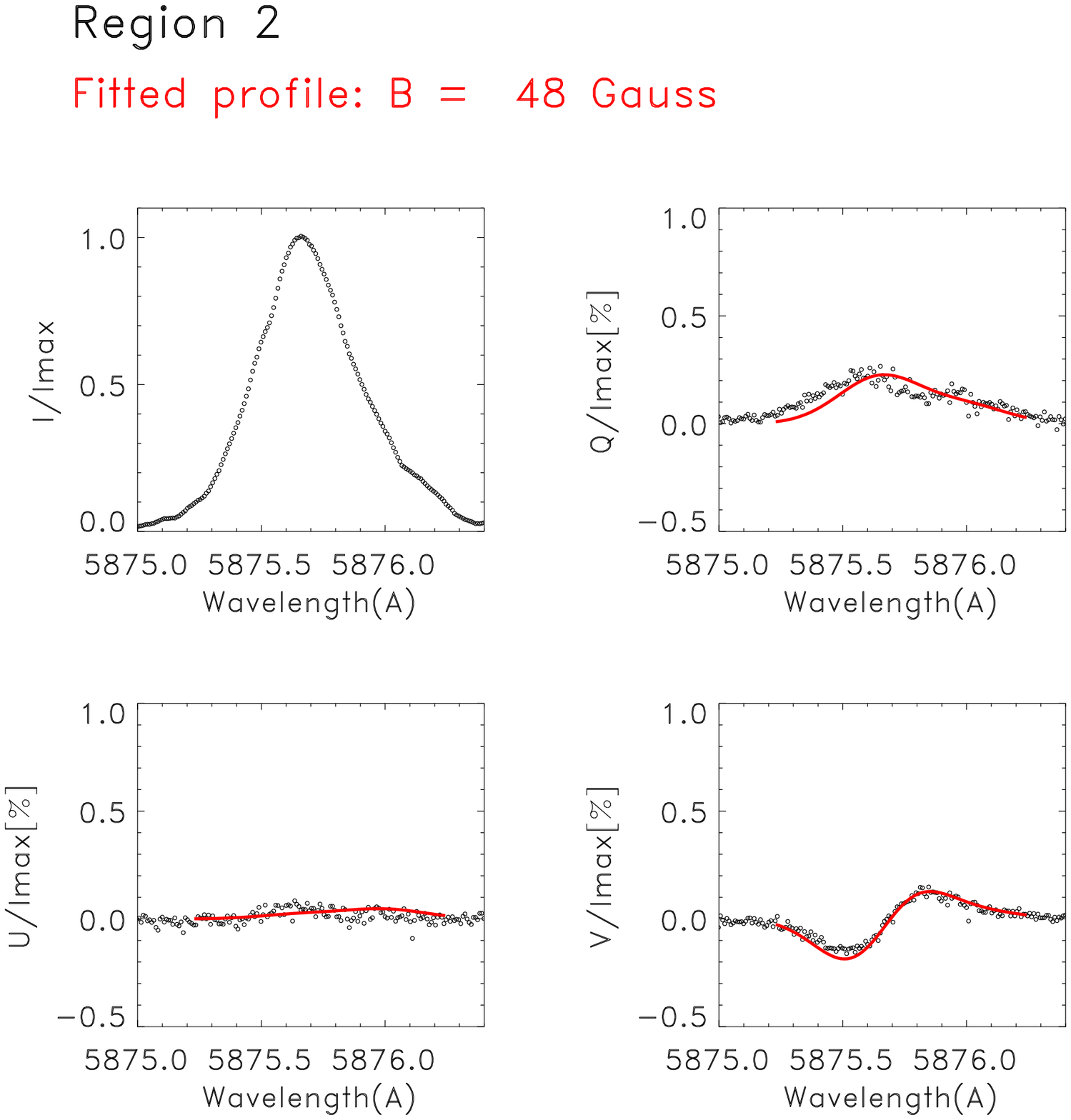}
\caption{\label{prof-act-71-79}Stokes profiles obtained after integrating 
the two regions selected in Fig.~\ref{5876-050618sp5876m4} together with fitted theoretical profiles (solid lines)
}
\end{figure}

Only in one measurement the
Stokes profiles indicate magnetic field strengths as high as
50-60 gauss (Figs.~\ref{5876-050618sp5876m4} and \ref{prof-act-71-79}). In this particular case the Stokes $V$ profile
shows a typical Zeeman-like antisymmetric shape. Note that
Stokes Q is affected by a strong depolarization. 
It is very important to point out that this 
measurement was obtained near the equator in the proximity of an
active region (AR 10776, 18 June 2005).

\section{Conclusion}
Using ZIMPOL at the GCT telescope in Locarno it was possible to obtain
a large set of high quality full Stokes
spectropolarimetric measurements  of spicules in He-D$_3$.
The observations of spicules in quiet regions
indicate that the magnetic fields involved
are around 10 gauss. In one measurement taken in the 
proximity of an active region, magnetic fields up to 50-60 gauss were found.

\acknowledgements 
We are grateful for the financial support that has been provided by
the canton of Ticino, the city of Locarno, ETH Zurich and the
Fondazione Carlo e Albina Cavargna.
This work has been also partially supported by the Spanish Ministerio de
Educaci\'on y Ciencia through project AYA2004-05792 and by the European
Solar Magnetism Network.

\end{document}